\documentclass[pdftex,preprint,twocolumn]{revtex4}
\usepackage[english]{babel}
\usepackage{times}
\usepackage{hyperref}
\usepackage[]{graphicx}

\begin{document}

\title{Behavior of some characteristics of EAS in the region of
knee and ankle of spectrum}

\author{S. P. Knurenko}\email[]{s.p.knurenko@ikfia.ysn.ru}
\author{A. V. Sabourov}\email[]{tema@ikfia.ysn.ru}
\author{I. Ye. Sleptsov}\email[]{i.ye.sleptsov.ikfia.ysn.ru}

\affiliation{Yu. G. Shafer Institute of Cosmophysical Research and
  Aeronomy, 31 Lenin Ave., 677980 Yakutsk, Russia}

\begin{abstract}
  The energy dependence of such characteristics as a ratio of
  the total number of charged particles to the total flux of EAS
  Cherenkov radiation, a ratio of $E_{thr} \ge 1$~GeV muon flux
  density at the distance of $600$~m from a shower core to charged
  particle flux density, a ratio of the energy transferred to the
  electromagnetic component of EAS to the primary particle energy is
  presented. Their comparison with two-component mass composition of
  cosmic rays (p-Fe) in the framework of calculations by a QGSJET
  model is given.
\end{abstract}

\maketitle

\section{Introduction}

The irregularities in the cosmic ray (CR) energy spectrum of ``knee''
type at $E_{0} \simeq 3 \times 10^{15}$~eV and ``ankle'' type at $E_{0}
\simeq 8 \times 10^{18}$~eV found in~\cite{ref1, ref2} are yet of
special interest from the point of view of interpretation of these
phenomena from the position of astrophysics. In recent years a few
papers~\cite{ref3, ref4, ref5} have been
published which try to explain such a behavior of CR spectrum with
the help of new models of generation and propagation of CRs. There
exists also another namely a nuclear-physical point of view for the
formation of irregularity at $E_{0} \simeq3 \times 10^{15}$~eV~\cite{ref6,
  ref7}. In some sense, the answer to the problem on
reasons of the formation of breaks is in the detailed study of
different EAS characteristics in the region of the first and second
irregularities in the spectrum. We have made such a work
in~\cite{ref8, ref9, ref10, ref11, ref12, ref13}.

\section{EAS characteristics in the superhigh energy region}

\subsection*{Longitudinal development}

\begin{figure}
\centering
\includegraphics[width=0.42\textwidth, clip]{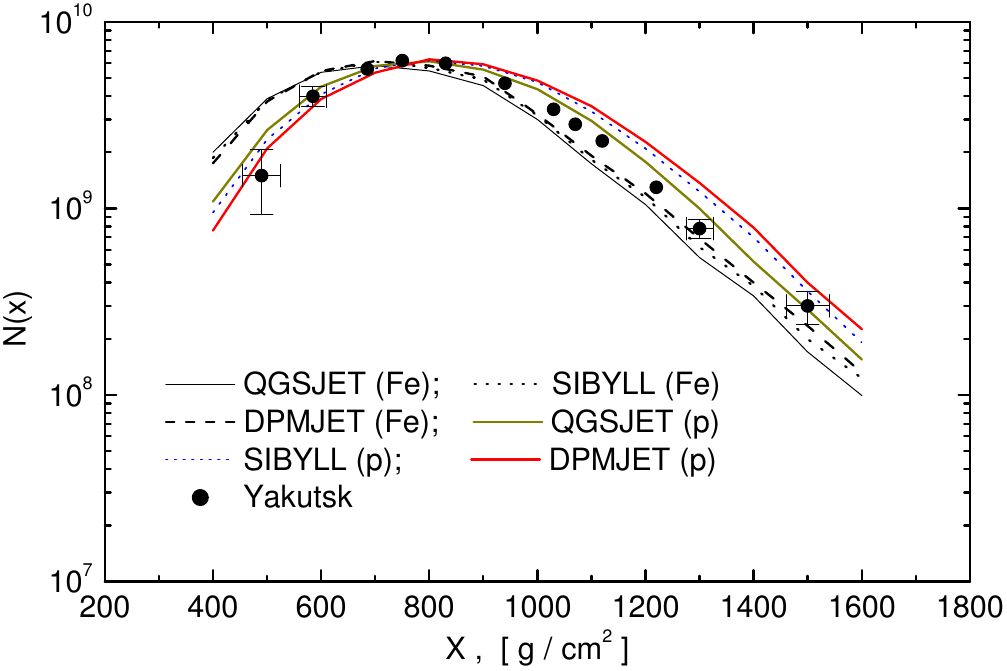}
\caption{Comparison of the experimental cascade curve of EAS
  development ($E_{0} = 10^{19}$~eV) with different models of hadron
  interactions.}
\label{fig2}
\end{figure}

The cascade curves of EAS development in 
Fig.\ref{fig2} were reconstructed according to the method suggested
in~\cite{ref14}. It is seen from 
Fig.\ref{fig2}
that the maximum depth of cascade curves depends on the primary
particle mass composition as well as the hadron interaction model. It
is seen from Fig.\ref{fig2} that to describe the experimental cascade
curve ($X_{max}$, $N_{0}$) the QGSJET model is
better-suited~\cite{ref15}. So we use this model for the estimation of
mass composition of primary particles. Fig.\ref{fig3} presents the
calculations of the maximum depth $X_{max}$ using the QGSJET model for
the primary proton and iron nucleus, and experimental data obtained at
the Yakutsk EAS array. It is seen that the velocity of shift of
$X_{max}$ to sea level depends on the energy range. In the framework
of the QGSJET model the experimental data are indicative of the change
of mass composition of primary particles in the energy range $E_{0} =
3 \times 10^{15} - 3 \times 10^{16}$~eV and at $E_{0} > 3 \times
10^{18}$~eV.


\subsection*{Radial development}

\begin{figure}
\centering
\includegraphics[width=0.42\textwidth, clip]{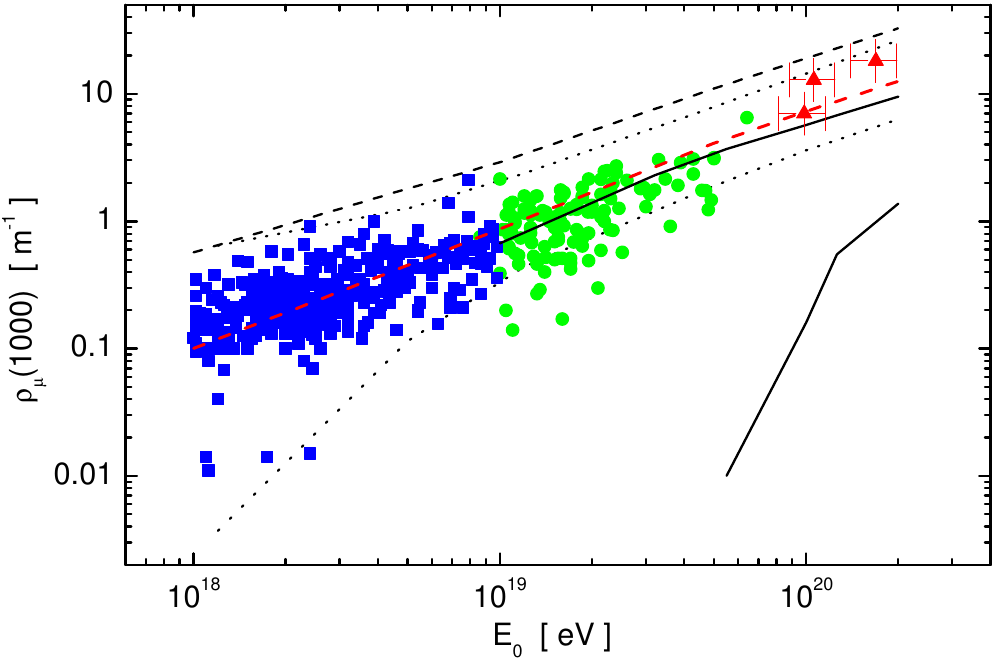}
\caption{$\rho_{\mu}(1000)$ vs $E_{0}$ relations for observed events
  $10^{18} - 10^{19}$~eV (squares), $10^{19} - 10^{20}$~eV (points)
  and $10^{20} - 10^{21}$~eV (triangles). Expected  $\pm 1\sigma$
  bounds for the distributions are indicated for p, Fe and $\gamma$
  primary by different curve as in the legend.}
\label{fig3}
\end{figure}

\begin{figure}
\centering
\includegraphics[width=0.42\textwidth, clip]{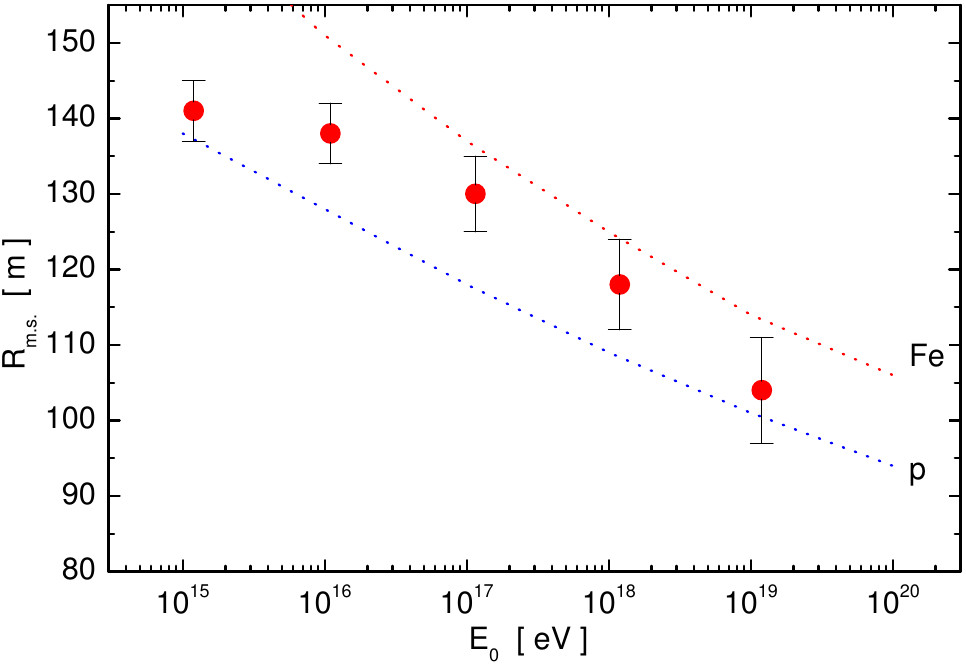}
\caption{Dependence of the root-mean-square LDF radius of charged
  particles on energy. The curves are calculations by the QGSJET model
  for the primary proton and iron nuclei.}
\label{fig4}
\end{figure}

Fig.\ref{fig3} and Fig.\ref{fig4} present experimental data: a) the
density of muons with $E_{thr} > 1$~GeV at a distance of 1000~m from
the shower core~\cite{ref11}, b) the root-mean-square radius $R_{m.s.}$
of charged particle (LDF)~\cite{ref12}. These data are compared with
calculations using the QGSJET model. Protons, iron nuclei and
$\gamma$-quanta are considered as primary particles. A confidence
interval taken in calculations is $\pm 1\sigma$. From Fig.3 it follows
that in the energy range $10^{18} - 5 \times 10^{18}$~eV within the
boundaries of confidence there are $\sim 60$\,\% of showers for the
iron nucleus and $\sim 90$\,\% for the proton. At $E_{0} > 5 \times
10^{18}$~eV the portion of iron nuclei decreases, the per cent of
protons and $\gamma$-quanta increases. It is seen from Fig.\ref{fig4}
that the ``heaviest'' mass composition is observed at $E_{0} \sim
10^{17}$~eV.

\subsection*{Correlation of EAS parameters}

\begin{figure}
\centering
\includegraphics[width=0.42\textwidth, clip]{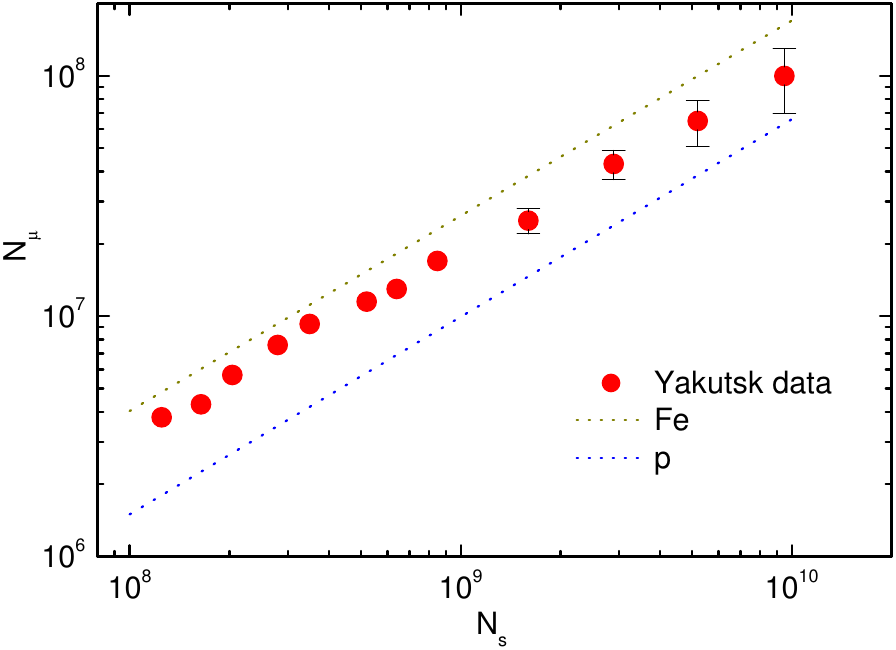}
\caption{Total number of charged particles $N_{s}$ and muons $N_{\mu}$
  at sea level. The curves are a calculation by the QGSJET model for
  the primary proton and iron nuclei.}
\label{fig5}
\end{figure}

\begin{figure}
\centering
\includegraphics[width=0.42\textwidth, clip]{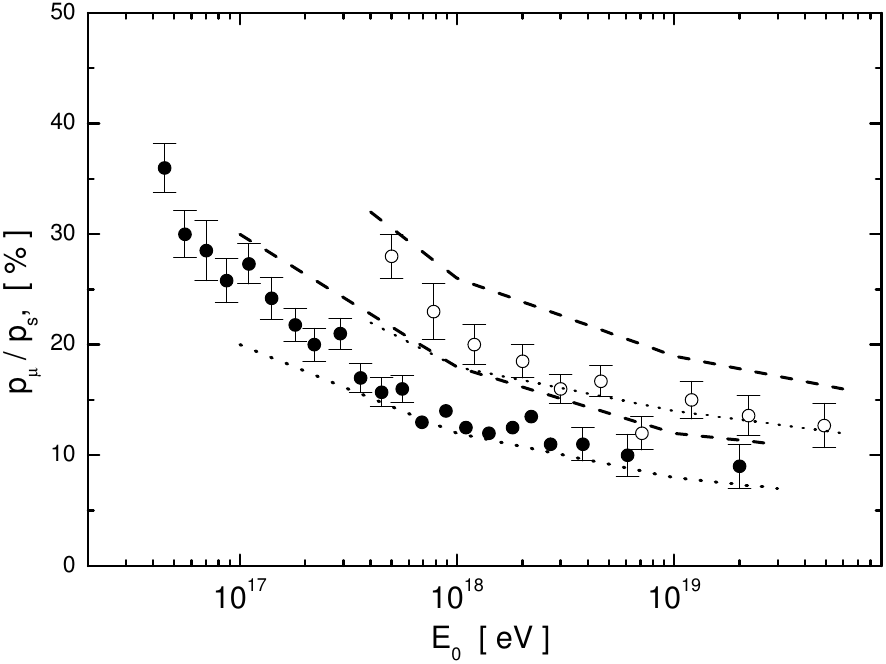}
\caption{Portion of the muons with $E_{thr} \ge 1$~GeV
  (\%). $\rho_{\mu}(300)/\rho_{s}(300)$ and
  $\rho_{\mu}(600)/\rho_{s}(600)$.}
\label{fig6}
\end{figure}

The most sensitive instrument for the model of hadron interactions and
mass composition of primary particles is the EAS muon component. At
the Yakutsk complex EAS array the muons with $E_{thr} \ge 1$~GeV are
measured by a shower registration in 70\,\% cases. Fig.\ref{fig5}
presents the correlation of $N_{\mu} - N_{s}$ parameters and
Fig.\ref{fig6} gives the portion of muons (ratio of muons to all
charged particles) at a distance of 300~m and 600~m from the shower
core. In the same place the calculations by the QGSJET model are
given. The tendency for an increase of light nuclei in the cosmic ray
primary flux at $E_{0} \ge 3 \times 10^{18}$~eV is marked by these
data.

\subsection*{Energetic EAS characteristics}

\begin{figure}
\centering
\includegraphics[width=0.42\textwidth, clip]{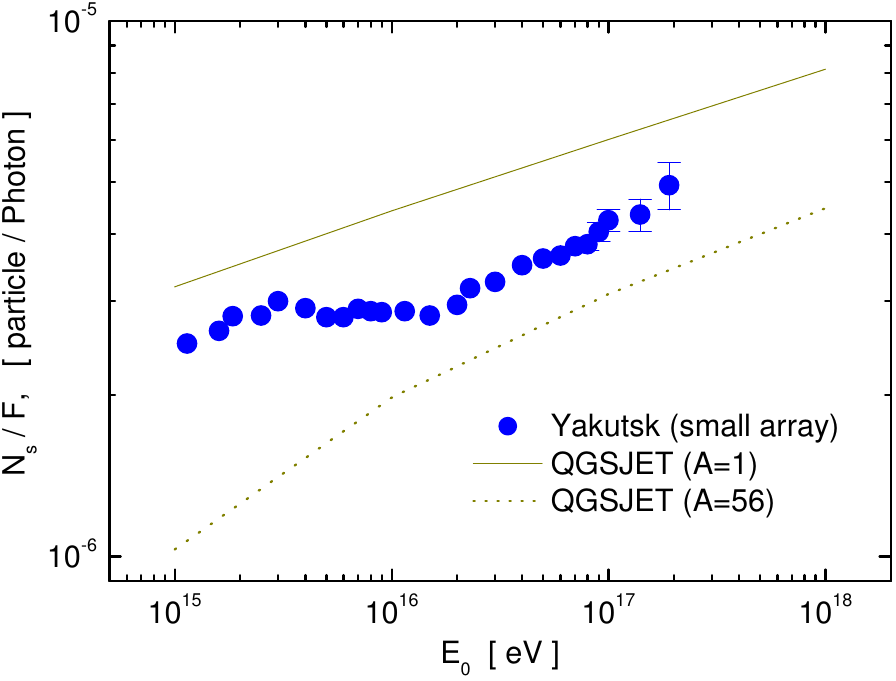}
\caption{The dependence of $N_{s} / F$ on a primary energy.}
\label{fig7}
\end{figure}

\begin{figure}
\centering
\includegraphics[width=0.42\textwidth, clip]{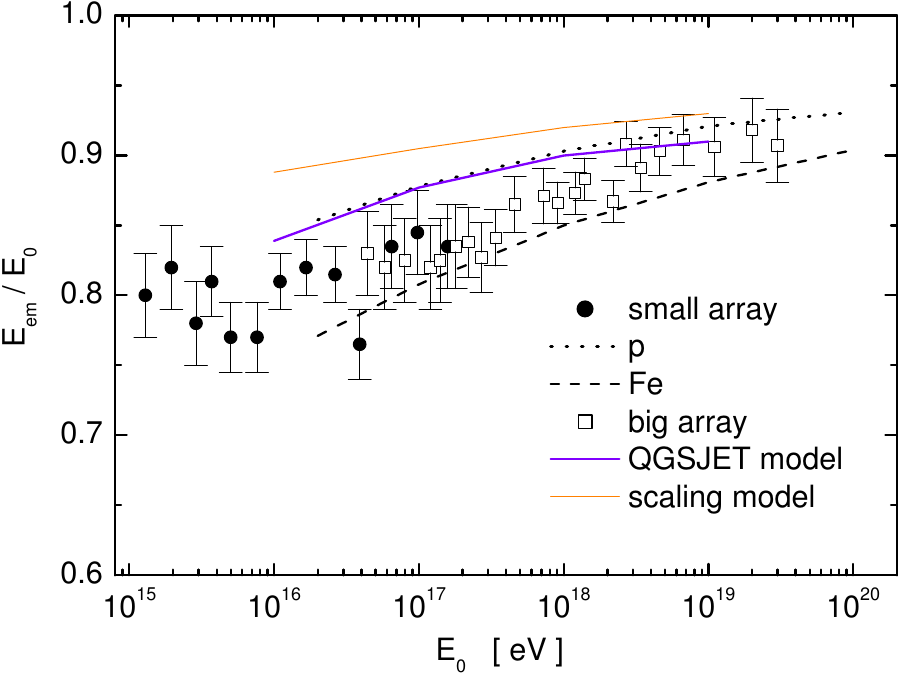}
\caption{A portion of the energy transferred to the electromagnetic
  EAS component  by Cherenkov light data at the Yakutsk array.}
\label{fig8}
\end{figure}

At the Yakutsk complex EAS array the shower energy is determined by
measurements of the total flux of EAS Cherenkov light $F$, the total
number of charged particles, $N_{s}$, and muons, $N_{\mu}$, at sea
level~\cite{ref16}. Fig.\ref{fig7} presents the energy-dependence of
$N_{s} / F$ ratio. It is seen from calculations that the ratio
strongly depends on a mass composition. From comparison of the
calculations by the QGSJET model for the proton, iron nucleus and
experimental data it follows that the mass composition changes just
after the first knee in the spectrum, i.e. in the interval $E_{0} = 5
\times 10^{15} - 10^{17}$~eV. It is evident from Fig.\ref{fig8} where
the portion of energy transmitted in to the electromagnetic EAS
cascade is shown.

\subsection*{Fluctuations of some EAS parameters}

\begin{figure}
\centering
\includegraphics[width=0.42\textwidth, clip]{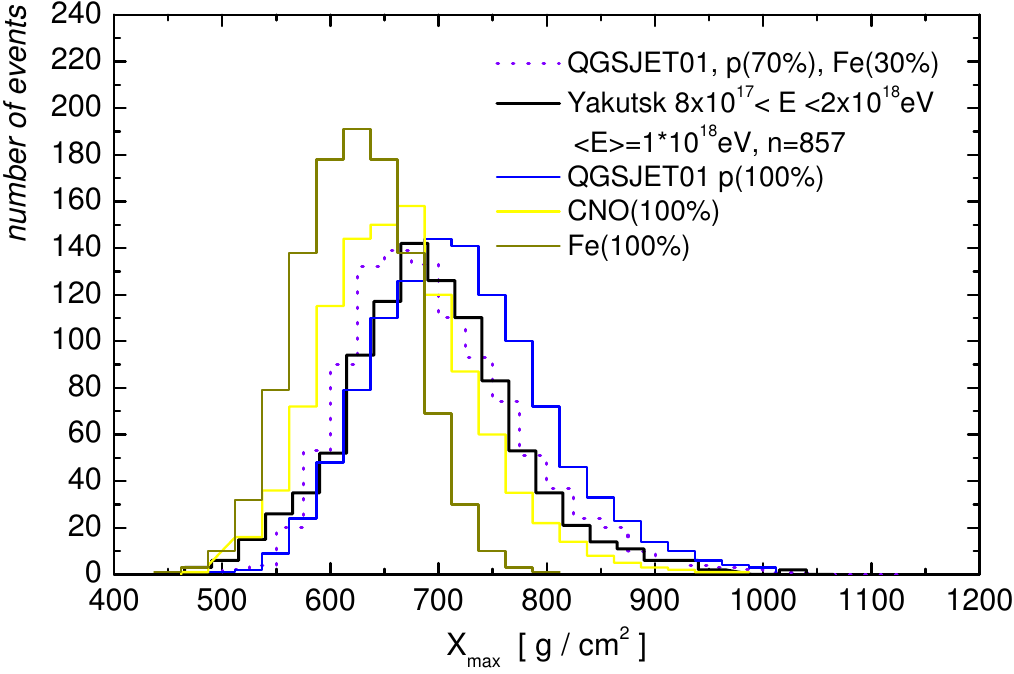}
\caption{Fluctuations of the parameter $X_{max}$ at $E_{0} \sim
  10^{18}$~eV.}
\label{fig9}
\end{figure}

\begin{figure}
\centering
\includegraphics[width=0.42\textwidth, clip]{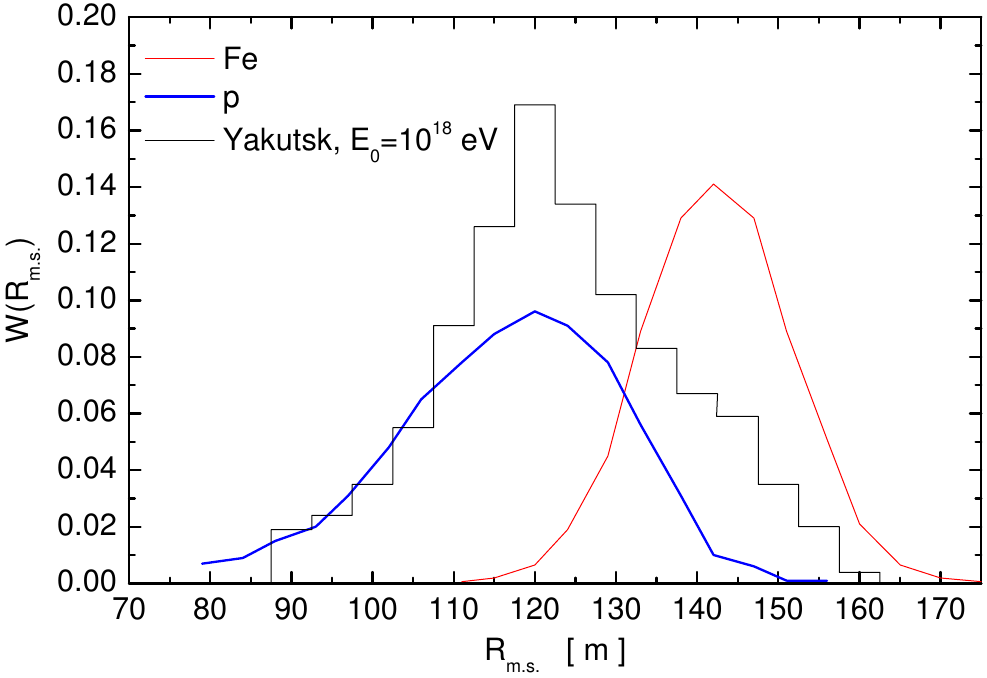}
\caption{Fluctuations of $R_{m.s.}$ at $E_{0} = 10^{18}$~eV. Curves
  are the QGSJET model calculations for the primary proton and iron
  nucleus.}
\label{fig10}
\end{figure}

In this section we consider the fluctuations of $X_{max}$ and
$R_{m.s.}$ obtained by the measurement of Cherenkov EAS light and
density of charged particle flux (see Fig.\ref{fig9} and
Fig.\ref{fig10}). In both cases at $E_{0} = 10^{18}$~eV fluctuations
are considerable and correspond to the mixed mass composition of
primary particles. From comparison with calculations by the QGSJET
model we have the following relationship: the light nuclei are $\sim
70$\,\% and heavy nuclei are $\sim 30$\,\% (see the dotted line in
Fig.\ref{fig9}).

\section{Results and Discussion}

At the Yakutsk complex  array for almost the 35-year period of
continuous observations the unique experimental data on electron, muon
and Cherenkov EAS components in the region of superhigh and ultrahigh
energies have been accumulated. Results on longitudinal and radial
development of EAS are presented in figures. From the total
combination of data one can select two energy regions where, as is
seen from Figures, the characteristics of EAS have the complex 
dependence on the energy. This is the energy ranges $10^{15} -
10^{17}$~eV and $10^{18} - 10^{19}$~eV. As is known, in these energy
ranges the irregularities of ``knee'' and ``ankle'' in the energy
spectrum of EAS are observed.

From the comparison of all experimental data with calculations by the
QGSJET model in~\cite{ref17} the results of the cosmic ray mass
composition in the energy range $10^{15} - 3 \times 10^{19}$~eV have
been obtained. It follows from the analysis that after the ``knee''
the mass composition becomes heavier and in the region of ``ankle'',
on the contrary, becomes lighter. Such a conclusion doesn't contradict
the hypothesis on a cosmic ray generation up to the energy $\sim
10^{18}$~eV in our Galaxy and their propagation according to the model
of anomalous diffusion in the fractal interstellar medium. Beginning
with $E_{0} > 3 \times 10^{18}$~eV the mass composition becomes
lighter and it doesn't contradict the presence of cosmic rays
metagalactic origin in the total flux of cosmic rays.

\end{document}